# Engineering the microstructure and magnetism of $La_2CoMnO_6$ thin films by tailoring oxygen stoichiometry

R. Galceran[1], C. Frontera[1], Ll. Balcells[1], J. Cisneros-Fernández[1], L. López-Mir[1], J. Roqueta[2], J. Santiso[2], N. Bagués[1,2], B. Bozzo[1], A. Pomar[1], F. Sandiumenge[1], B. Martínez[1]

[1]Institut de Ciència de Materials de Barcelona-CSIC, Campus UAB, E-08193 Bellaterra, Spain

[2]Institut Català de Nanociència i Nanotecnologia, ICN2-CSIC, Campus UAB, E-08193 Bellaterra, Spain

**ABSTRACT**

We report on the magnetic and structural properties of ferromagnetic-insulating $La_2CoMnO_6$ thin films grown on top of (001) STO substrates by means of RF sputtering technique. Careful structural analysis, by using synchrotron X-ray diffraction, allows identifying two different crystallographic orientations that are closely related to oxygen stoichiometry and to the features (coercive fields and remanence) of the hysteresis loops. Both Curie temperature and magnetic hysteresis turn out to be dependent on the oxygen stoichiometry. In situ annealing conditions allow tailoring the oxygen content of the films, therefore controlling their microstructure and magnetic properties.

Bulk La$_2$CoMnO$_6$ (LCMO) double perovskite has been the subject of strong interest during the last years [1, 2, 3, 4, 5, 6, 7] due to its ferromagnetic insulating character. Besides its perspectives as magnetodielectric material, LCMO is a good candidate for active insulating barriers in spin filters. For these devices, insulating barriers must be thin enough to enable tunneling conduction. The properties of epitaxial LCMO thin films have been the subject of some theoretical and experimental works in the recent years [8, 9, 10, 11]. Previous experimental reports based on films prepared by pulsed laser deposition (PLD) suggest that magnetic properties are strongly dependent on growth conditions. When samples are grown under low oxygen pressure the Curie temperature ($T_C$) is around 100K while increasing oxygen pressure (200 mTorr and above) $T_C$ values around 230 K can be achieved. Nevertheless, there is no clear consensus on whether this variation of $T_C$ comes from differences on the Co/Mn cationic ordering [10] or from changes in the oxygen stoichiometry [9]. On the other hand, low temperature hysteresis loops reported in the literature present anomalies and "biloop" features that have been attributed to the existence of a bidomain structure in the films. However, these "biloop" features of hysteresis cycles are present up to $T_C$=230K and therefore, cannot be linked to the persistence of regions with low $T_C$ phase [10].

Previous studies report low temperature hysteresis loops with a saturation magnetization close to 6 $\mu_B$/f.u. [8, 9, 10], i.e., the expected value of spin-only saturation magnetization in the case of ferromagnetic (FM) ordering of Mn$^{4+}$ and Co$^{2+}$ (in high spin state) ions. The existence of disorder in the cation's occupancy of the double perovskite structure, namely the presence of Mn ions placed in Co sublattice and *vice versa* (antisites), will introduce $t_{2g}$-$t_{2g}$ (Mn$^{4+}$-Mn$^{4+}$) and $e_g$-$e_g$ (Co$^{2+}$-Co$^{2+}$) superexchange interactions that according to Goodenough-Kanamori-Anderson rules [ 12, 13, 14] are antiferromagnetic (AF). Therefore, it is expected that the existence of antisites will reduce the saturation magnetization from 6 $\mu_B$/f.u.. Thus, the departure of the saturation magnetization of a given sample from this saturation value can be interpreted as a measure of the degree of Co/Mn disorder in the structure.

The effect of Co/Mn disorder on the magnetic properties has been previously studied in bulk samples (where it can be precisely quantified by means of neutron powder diffraction) [ 6]. These studies show that hysteresis loops become wider (with higher coercive field and lower remanence) in the presence of disorder [ 2, 4, 6]. Co/Mn ordering temperature is around 1125ºC and the ordering process is blocked below 1000ºC due to extremely large relaxation times [ 3]. The reported valences for Co and Mn are 2+ and 4+ respectively and barely change with the degree of cationic ordering [ 3, 6].

In this work we have studied the growth conditions of high quality LCMO epitaxial thin films to explore their suitability as spin filters. Samples have been prepared on top of (001) oriented $SrTiO_3$ (STO) substrates by means of RF magnetron sputtering. The target was prepared by solid state reaction according to the details published elsewhere [ 6]. Films have been grown at heater temperatures between 800ºC and 900ºC (substrate temperature must be slightly below heater temperature) with partial oxygen pressure ranging from 200 to 400 mTorr and with different in situ thermal treatments after deposition. Sample names and preparation conditions are listed in Table I. Structural properties of thin films were studied by X-ray diffraction and reflectivity using a Rigaku Rotaflex RU200B diffractometer equipped with a rotatory anode. The thickness of LCMO films, determined by X-ray reflectivity, depends on the oxygen pressure and spans from 35nm (200 mTorr) to 20 nm (400 mTorr). Additionally, some selected samples were analyzed by means of synchrotron radiation at KMC2 beamline (BESSY-II at Helmholtz-Zentrum, Berlin), which is equipped with a two dimensional General Area Detector Diffraction System (GADDS), for a deeper structural characterization. Magnetization measurements were performed using a superconducting quantum interference device (SQUID, Quantum Design) as a function of temperature and magnetic field applied along $(100)_{STO}$ crystallographic direction. Transport measurements were carried out in a Physical properties Measurement System (PPMS, Quantum Design) by using the standard four probe method. Topography of films was investigated by atomic force microscopy (AFM), using an Asylum Research MFP-3D microscope in tapping mode, revealing a flat surface with terraces-and-steps morphology which follows the STO surface morphology (see inset in Fig. 2).

Fig. 1 shows the magnetization *vs*. temperature curves, measured following a zero-field-cooling field-cooling process under H= 1kOe, for the most representative samples. Top panel illustrates the dependence of magnetic properties on growth temperature ($T_g$): increasing $T_g$ from 800ºC to 900ºC promotes an increase of $T_C$ from ~130K to ~160K. The dependence of $T_C$ on the oxygen pressure ($p_{O2}$) at a fixed $T_g$ (=900ºC) is evidenced in Fig. 1b, which reveals that lowering $p_{O2}$ from 400 to 300 mTorr promotes the reduction of $T_C$ down to ~120K.

To further enhance $T_C$, we have performed different in-situ annealing processes and depicted the results in Fig. 1c. This figure shows that by annealing at 900 ºC and $p_{O2}$= 400 Torr we obtain $T_C$ values of about 225 K, close to that reported for bulk samples [ 2, 4] and films grown by PLD [ 8, 9, 10]. However, we note that PLD films were grown at 800 ºC and no annealing process was needed, whilst for RF sputtered films $T_g$=800 ºC is not enough to obtain a high $T_C$, even after an *in-situ* annealing at $p_{O2}$= 400 Torr (sample B, not shown). Furthermore, we have evaluated the influence of lowering the cooling rate after the annealing (sample F), observing a slight increase of Tc (~230 K). We have measured the temperature dependence of the resitivity of different samples, all of which display insulating behavior. As an example, resitivity of sample F is displayed in the inset of Fig. 2.

Fig. 2 plots the M(H) hysteresis loops corresponding to samples with low and high $T_C$ (C, E, and F) measured at T=10K. The diamagnetic contribution of the substrate has been corrected by subtracting the corresponding M(H) loops measured at 350K. In the case of sample C (low $T_C$), we find a coercive field of $H_C$~5200 Oe, with saturation magnetization $M_S$~5.0 $\mu_B$/f.u. at H= 70 kOe. For sample F (high $T_C$) a loop with a higher value of the coercive field ($H_C$~7500Oe) is found but irreversibility extends up to very high fields (~50 kOe). In fact magnetization attains a value near 4.33 $\mu_B$/f.u. (at H= 70 kOe) but it is still far from reaching the saturation regime. In the case of sample E ($T_C$~225 K, slightly below that of sample F), which was annealed in the same conditions as sample F but cooled down at a faster cooling rate, coercive field is $H_C$~7100 Oe and magnetization does not saturate at the highest attainable field. Interestingly, this sample presents a biloop that mimics a superposition of loops of samples C and F. This double loop shape has been previously reported and considered as evidence of the existence of a bidomain structure [ 10].

To clarify the possible relation between the magnetic behavior and the structure of the samples we have carefully analyzed samples C, D and F by means of synchrotron diffraction. First we have examined the reciprocal space maps of the (103) diffraction peak (indexed with respect to STO). These maps (not shown) reveal that the films are epitaxial with the same in-plane lattice parameters as the STO substrate. Therefore, no relaxation of the tensile strain due to lattice mismatch with the substrate occurs during annealing processes. Rods along $Q_Z$ show thickness fringes revealing the high crystal quality of the prepared films. To elucidate if LCMO grows with *c*-axis parallel or perpendicular to the substrate surface (sketch at the top of Fig. 3), we measured the $(0\bar{3}\frac{5}{2})$ and $(0\frac{\bar{5}}{2}3)$ diffraction peaks (referred to STO reciprocal lattice) which appear exclusively when the *c*-axis of LCMO is out-of-plane or in-plane, respectively. We performed phi-scans around their positions in reciprocal space and plot the projections along $Q_x$ of the collected intensity in Fig. 3. The orientation of c-axis strongly varies with preparation conditions. For sample C, reflection $(0\bar{3}\frac{5}{2})$ is much stronger than the $(0\frac{\bar{5}}{2}3)$ one (see Fig. 3a and 3b), proving that both orientations coexist but that the *c*-in plane orientation is predominant. For sample E both peaks present similar intensities, thus indicating that both orientations coexist in similar proportions, in clear relation with the presence of a biloop in the M(H) curve. Finally, sample F presents only the $(0\frac{\bar{5}}{2}3)$ reflection, representative of a unique crystallographic orientation (*c*-out of plane). In accordance, the hysteresis cycle does not show any evidence of a double coercive field. For the phase with *c*-out of plane, the *c* lattice parameter decreases monotonically with the fraction of that phase: 7.803(5), 7.762(5) and 7.736(5) Å for samples C, E and F respectively. However, this change of *c* parameter cannot be attributed to a relaxation of structural strain, as mentioned above.

The differences in the growth conditions of the studied samples are the thermal treatment and the oxygen partial pressure in the chamber. In order to discern which of these two factors determines the change in $T_C$, we have prepared a new sample (G) that was annealed at 900ºC during 1 hour but maintaining a low $p_{O2}$ pressure (400 mTorr, as during the growth process), in contrast to the 400 Torr used for samples E and F. This sample exhibits low $T_C$ (~150K) very similar to that of sample D, even when its thermal history is identical to sample E, but with a reduced $p_{O2}$ pressure during the annealing. Therefore, the increase of the oxygen content is the main responsible for the large increase in $T_C$. This result is in accordance with previous studies in bulk $La_2CoMnO_{6-\delta}$ samples (Ref. 2) where low $T_C$ (~150K) was reported for samples with $\delta \geq 0.05(1)$ but high $T_C$ (~225K) was found when the oxygen deficiency is $\delta \leq 0.02(1)$; and also in agreement with the conclusions derived in Ref. [ 9] for thin films grown by PLD, where the change in $T_C$ is attributed to variations of oxygen stoichiometry. However, the annealing temperature also plays a relevant role. As shown in the case of sample B, an annealing process at 800ºC is not enough to increase $T_C$, which indicates low oxygen content. This result proves that oxygen diffusion proceeds slowly at 800ºC but much faster at 900ºC. Hence, a threshold temperature, between these two values, must be surpassed to reach an efficient oxygen uptake and diffusion inside the films. The behavior shown by sample F (slowly cooled from 900ºC) can be explained when considering that the slow cooling process implies a longer time above this threshold temperature, resulting in a larger oxygen uptake. In fact, samples with prolonged annealing time (two hours or more) and fast cooling exhibit magnetic behavior similar to that of sample F.

Consequently, there exists a correlation between oxygen stoichiometry and the two different crystallographic orientations: on increasing the oxygen content, the amount of phase with *c*-out of plane increases, as evidenced by our synchrotron study. The variation of the *c*-cell parameter (of the phase with *c*-out of plane) also supports the idea of a progressive increase of the oxygen content. The variation of the cell volume (59.494, 59.181, 58.983 Å$^3$) for samples C, E and F respectively also tends to that reported for the bulk value (58.771 Å$^3$ [ 6]). This shrinkage of the *c*-parameter (and cell volume) with the oxygen content is coherent with the reduction of the cationic radii under oxidation (ionic radii of Mn$^{3+}$ and Mn$^{4+}$ are 0.785 and 0.670 Å respectively [ 15, 16]).

As mentioned previously, cationic ordering is a key parameter for controlling the magnetic properties of bulk LCMO [ 3, 4, 6]. In previous works on thin films prepared by PLD, cationic ordering was inferred in base of $M_S$ values at low T close to 6 $\mu_B$/f.u., the expected value for spin-only fully-ordered La$_2$Co$^{2+}$Mn$^{4+}$O$_6$ ferromagnet [ 8, 9, 10]. Our study evidences that magnetization at high fields depends on the oxygen stoichiometry. Sample C, with low oxygen content, saturates clearly below 6 $\mu_B$/f.u., thus it presents some degree of Co/Mn disorder. For films with high $T_C$, i.e. those with high oxygen content, M(H) curves do not reach saturation, therefore, we cannot establish the degree of order following this criterion.

At this point it is important to recall that similar $T_C$ values were found in bulk samples with different degree of Co/Mn cationic ordering [ 2, 6]. However, bulk samples with high degree of Co/Mn ordering exhibit hysteresis loops with low $H_C$ values and high remanence, whilst samples with low Co/Mn ordering present high $H_C$ values and low remanence [ 2, 4, 6]. Consequently, the high $H_C$ and low remanence of samples E and F indicate that the degree of Co/Mn ordering is not optimal either. This result is congruent with the fact that, in LCMO bulk material, the Co/Mn ordering process is blocked below 1000ºC [3], so no enhancement of cationic order can be expected during annealing processes at 900 ºC.

In conclusion, we have prepared epitaxial thin films (~20 nm) of the ferromagnetic insulating LCMO compound by RF magnetron sputtering with high $T_C$ values. Synchrotron diffraction studies show that films grow epitaxially on top of (001) STO substrate and are fully strained and of high crystalline quality. A strong dependence of $T_C$ on the oxygen content was detected. Only samples annealed at high enough temperature (~900ºC) in oxygen rich atmosphere ($p_{O2}$=400 Torr) exhibit $T_C$ values well above 200 K. On the other hand, we have evidenced that oxygen content also controls the abundance of the two different crystallographic orientations existing in the films; high oxygen content (sample F) leads to a single crystallographic orientation (with *c*-axis out of plane), while for low oxygen content both crystallographic orientations coexist. However, magnetization curves at low T indicate that Co/Mn cations are not well ordered.

Although at this point we have achieved ferromagnetic, insulating, flat-surface thin (~20nm) films of high crystalline quality, further studies are required in order to ensure good capabilities of LCMO films as spin filter barriers.

**ACKNOWLEDGEMENTS**

We acknowledge financial support from the Spanish MEC (MAT2011-29081 and MAT2012-33207), CONSOLIDER (CSD2007-00041), and FEDER program. R.G., L.L.-M. and N.B. thank the Spanish MINECO for the financial support through the FPI program. We thank Helmholtz-Zentrum Berlin for the allocation of synchrotron radiation beamtime, and Dr. Daniel M. Többens for his kind assistance during data collection. The research leading to these results has received funding from the European Community's Seventh Framework Programme (FP7/2007-2013) under Grant Agreement No. 312284.

**TABLES**

| Table I: Growth conditions of the $La_2CoMnO_6$ films prepared by RF magnetron sputtering. | | | | |
|---|---|---|---|---|
| Name | $T_g$ (ºC) | $p_{O2}$ (mTorr) | $T_C$ (K) | In situ thermal treatment (after deposition) |
| A | 800 | 400 | 130 | Fast Cooling (25ºC/min, $p_{O2}$=400 mTorr) |
| B | 800 | 400 | 140 | Annealing 1h@800ºC, $p_{O2}$=400 Torr; cooling 10ºC/min |
| C | 900 | 300 | 120 | Fast Cooling (25ºC/min, $p_{O2}$=300 mTorr) |
| D | 900 | 400 | 160 | Fast Cooling (25ºC/min, $p_{O2}$=400 mTorr) |
| E | 900 | 400 | 225 | Annealing 1h@900ºC, $p_{O2}$=400 Torr; cooling 10ºC/min |
| F | 900 | 400 | 230 | Annealing 1h@900ºC, $p_{O2}$=400 Torr; cooling 1ºC/min<br>Annealing 4h@600ºC, $p_{O2}$=400 Torr; cooling 10ºC/min |
| G | 900 | 400 | 150 | Annealing 1h@900ºC, $p_{O2}$=400 mTorr; cooling 10ºC/min |

**FIGURE LEGENDS**

Fig. 1 Normalized magnetization vs. temperature of different films, measured under 1kOe. (a) Effect of growth temperature; (b) effect of oxygen partial pressure and (c) effect of annealing at 400Torr and effect of lowering cooling rate.

Fig. 2 Magnetization vs. applied field measured at 10K with H applied along (100)STO crystallographic orientation. Curves have been corrected for the diamagnetic contribution from the STO substrate. Insets show the AFM image (top left) and resistivity dependence on temperature (bottom right) corresponding to sample F.

Fig. 3 Reciprocal space maps measured by synchrotron X-ray diffraction for samples with the lowest and higher $T_C$; (right column) $(0\bar{3}\tfrac{5}{2})$ reflection, only present for $La_2CoMnO_6$ with $c$-out of plane, (left column) $(0\tfrac{\bar{5}}{2}3)$ reflection, only present for $La_2CoMnO_6$ with $c$-in plane. The sketch at the top of the figure illustrates the two possible orientations of the film on top of STO.

1 **FIGURES**

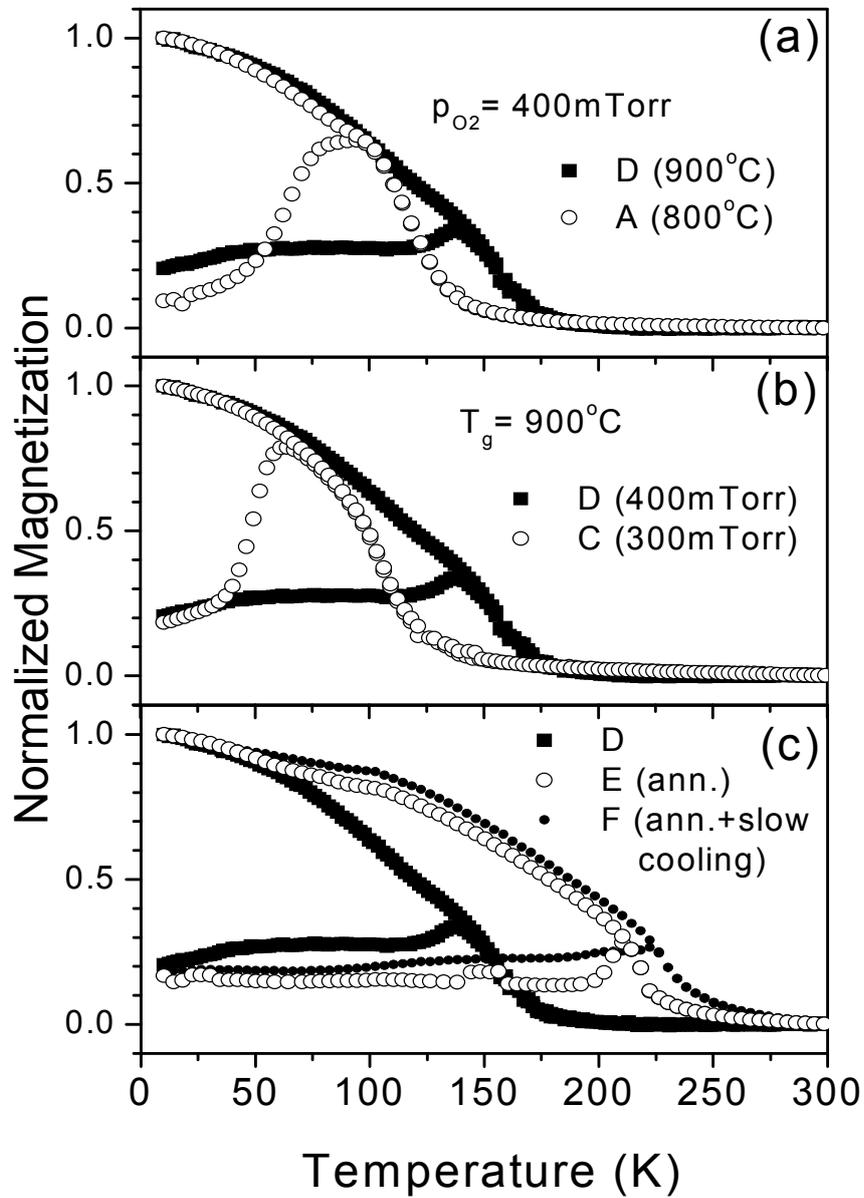

2
3

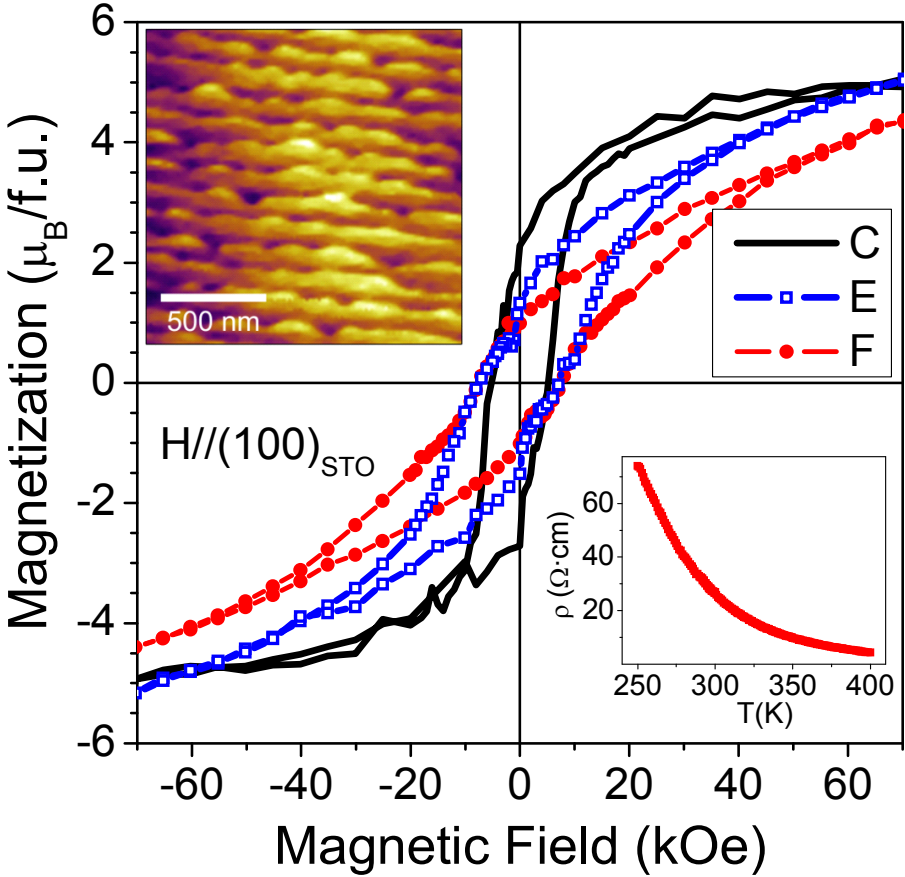

1
2

Engineering the microstructure and magnetism of $La_2CoMnO_6$ thin films by tailoring oxygen stoichiometry

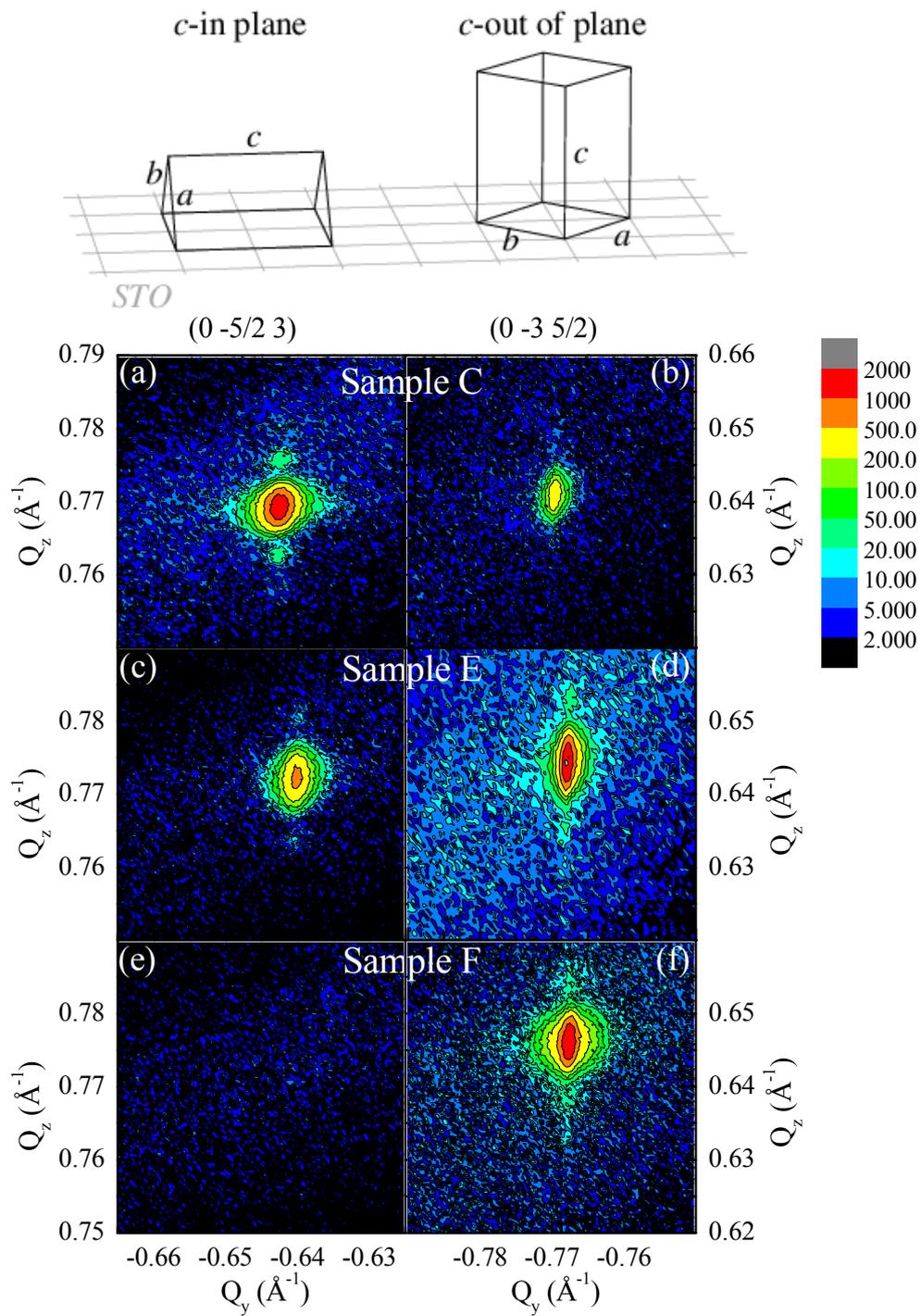

Engineering the microstructure and magnetism of $La_2CoMnO_6$ thin films by tailoring oxygen stoichiometry